\documentclass[preprintnumbers,amsmath,amssymb,floatfix,9pt,prd,
onecolumn,showpacs,superscriptaddress,nofootinbib]{revtex4}
\usepackage{graphicx}
\usepackage{epsfig}
\usepackage{bm}
\usepackage{amsfonts}
\usepackage{hyperref}

\usepackage{color}
\begin{document}

\title{Geometrothermodynamics of Myers-Perry black holes  }

\author{\textbf{ Alessandro Bravetti}} \email{bravetti@icranet.org}
\affiliation{Dipartimento di Fisica and ICRA, "Sapienza" Universit\`a di Roma, Piazzale Aldo Moro 5, I-00185 Rome, Italy}\affiliation{Instituto de Ciencias Nucleares, Universidad Nacional Aut\'onoma de M\'exico, AP 70543, M\'exico DF 04510, Mexico}
\author{\textbf{Davood Momeni}}
\email{momeni_d@enu.kz }
 \affiliation{Eurasian International Center for Theoretical Physics,  L.N. Gumilyov Eurasian National University, Astana 010008, Kazakhstan}
\author{\textbf{ Ratbay Myrzakulov}}
\email{rmyrzakulov@gmail.com}\affiliation{Eurasian International
Center for Theoretical Physics,  L.N. Gumilyov Eurasian National
University, Astana 010008, Kazakhstan}
\author{\textbf{ Aziza Altaibayeva}}
\email{aziza.ltaibayeva@gmail.com}\affiliation{Eurasian International
Center for Theoretical Physics,  L.N. Gumilyov Eurasian National
University, Astana 010008, Kazakhstan}

\begin{abstract}
{\bf Abstract:} 
We consider the thermodynamics and Geometrothermodynamics of the Myers-Perry black holes in five dimensions for three different cases, depending on the 
values of the angular momenta. We follow Davies approach to study the thermodynamics of black holes and find a non-trivial thermodynamic structure in all cases, which
is fully reproduced by the analysis performed with the techniques of Geometrothermodynamics.  
Moreover, we observe that in the cases when only one angular momentum is present or the two angular momenta are fixed to be equal, i.e. when the thermodynamic system is two dimensional, there is
a complete agreement between the divergences of the generalized susceptibilities and the singularities of the equilibrium manifold, whereas when the two angular momenta 
are fully independent, that is, when the thermodynamic system is three dimensional, additional singularities in the curvature appear. However, we prove that
such singularities are due to the changing from a stable phase
to an unstable one.
\end{abstract}
\pacs{04.50.+h, 04.70.Dy} 
\maketitle
\newpage
\section{Introduction}

Black holes are very special thermodynamic systems. They are thermodynamic system since they have a temperature, the celebrated Hawking temperature \cite{Hawking},
and a definition of entropy via the Bekenstein area law \cite{Bekenstein},  from which one can prove that the laws of thermodynamics apply to black holes \cite{Bardeen}.
On the other side they  are very special thermodynamic systems, since for instance the entropy is not extensive,  
they cannot be separated into small subsystems and, 
perhaps the worst fact, their thermodynamics does not possess a microscopic description yet (see e.g. \cite{ArcioniTalla} for a clear description of these problems).

In this puzzling situation, one of the most successful and at the same time discussed approach to the study of black holes phase transitions is the work of Davies \cite{Davies}.
According to Davies, black holes can be regarded as ordinary systems, showing phase transitions right at those points where the generalized susceptibilities, i.e. second
order derivatives of the potential, change sign most notably through an infinite discontinuity.
Since there is no statistical mechanical description of black holes as thermodynamic systems, it is hard to verify Davies approach with the usual technique 
of calculating the corresponding critical exponents (although very interesting works on this subject exist, see for instance \cite{Lousto, Lousto2, cai1,cai2, Bana1, Bana1bis, Bana1ter, Bana1quater, Bana2, 
Bana2bis, Bana2ter, Bana2quater, ArcioniTalla}).
In fact, the main drawback of this approach is that one has to choose arbitrarily  the order parameter for black holes.

A possible resolution to this situation can then come from the use of thermodynamic geometry.
Since the pioneering works of Gibbs \cite{Gibbs} and Carath\'eodory \cite{cara}, techniques form geometry have been introduced into the analysis of thermodynamics.
In particular, Fisher \cite{fisher} and Rao \cite{rao} proposed a metric structure in the space of  probability distributions which has been extensively used both in statistical physics 
and in economics (for a recent review see \cite{brody}).
Later, Weinhold \cite{wein} introduced an inner  product in the equilibrium space of thermodynamics based on the stability conditions of the internal energy, taken as the 
thermodynamic potential. The work of Weinhold was then developed by Ruppeiner \cite{rupp1} from a different perspective. Ruppeiner moved from the analysis of fluctuations
around equilibrium and from the gaussian approximation of the probability of fluctuations and found a thermodynamic metric  which is defined as (minus) the hessian
of the entropy of the system.  Remarkably, Ruppeiner geometry was found to be conformally related to the one proposed by Weinhold.
Moreover, Ruppeiner metric is intrinsically related to the underlying statistical model, in fact the scalar curvature of the Riemannian manifold representing the system using
Ruppeiner metric happens to have exactly the same exponent as the correlation volume for ordinary systems (see e.g. \cite{rupp2} for a review).

All these approaches have been widely used to study ordinary systems and in particular Ruppeiner metric has been also used to study many black holes configurations (see \cite{rupp3} and 
refences therein).
This is because one can argue that, being Ruppeiner metric defined only from thermodynamic quantities and on the other side giving informations about the statistical model, then it can provide 
some hints towards the resolution of the long standing problem of understanding the microscopic properties of black holes (see e.g. \cite{rupp3}).

On the other side, the problem with the use of thermodynamic geometries to study black holes thermodynamics is that black holes are not ordinary systems, as we argued above.
For instance, Ruppeiner metric in many cases gives exactly the same results of Davies approach (which is based upon ordinary thermodynamics),
while in some other important cases it does not converge to the same results, as it happens for example in the Reissner-Nordstr\"om and Kerr cases (see e.g. \cite{rupp3,aman1,aman2}).
One can argue either that Davies approach is inaccurate, or that the application of Ruppeiner metric to black holes may be imperfect, 
due to the strange nature of black holes as thermodynamic systems. 
In fact there is still an open debate on this topic (see e.g. the
discussion in \cite{rupp3,laszlo}).

Furthermore, in the recent years, a new approach in the context of thermodynamic geometry has been proposed by Quevedo  \cite{quev07}, known as 
Geometrothermodynamics (GTD).
According to this approach, the Riemannian structure to be introduced in the equilibrium space should be consistent with the property of Legendre invariance, a property
which is at the core of ordinary thermodynamics.
In GTD some families of metrics have been found that share the Legendre invariance property and they have also been proven to interpret in a consistent manner the thermodynamic
properties of ordinary systems, chemical interactions,  black holes configurations and cosmological solutions (see \cite{quev07, hernando2, PhTransGTD, quev08, KSBH, quevdiego,HanChen,AlejQuev,Brasil}).
In particular, the correspondence between the divergences of the scalar curvature of the equilibrium manifold of GTD and the phase transition points signaled by the divergences 
of the heat capacity (i.e. phase transitions {\it \`a la Davies}) seems to be a general fact, 
according to the variety of systems analyzed so far and to the general expressions given in \cite{Termometrica}.
In addition, a recent study \cite{AleQuevDavood} of the thermodynamics of the Reissner-Nordstr\"om and Kerr black holes in any dimensions suggested that 
the GTD approach can detect
not only the points of phase 
transitions due to singularities of the heat capacities, but also divergences of the full spectrum of the generalized susceptibilities.

On the other side, the thermodynamic properties of the Myers-Perry black holes in five dimensions have been extensively studied in the literature from completely different
points of view  (see e.g. \cite{ArcioniTalla,aman2, aman2bis, EmpMyers,MonteiroPerry,AsteMann,AsteRodriguez}).
In this work, we give special emphasis on the relation between divergences
of the generalized susceptibilities and curvature singularities of the metric from GTD.
For example, we do not consider here possible phase transitions related to change in the topology of the event horizon, an intriguing question which was addressed e.g. in
 \cite{ArcioniTalla}.
We find out that the GTD thermodynamic geometry is always curved for the considered cases, 
showing the presence of thermodynamic interaction, and that its singularities always correspond to divergences of the susceptibilities or to points where there is a change from a stable to an unstable phase.
This will allow us to infer new results on the physical meaning of the equilibrium
manifold of GTD.

The work is organized as follows. In Sec. \ref{sec:gtd} we present the basic aspects of GTD and introduce all the mathematical concepts that are needed. 
In Sec \ref{sec:MP} we perform the parallel between the thermodynamic quantities and the Geometrothermodynamic description of the five dimensional Myers-Perry black holes
for three different cases, depending on the values of the angular momenta.
Finally, in Sec. \ref{sec:con} we comment on the results and discuss possible developments.

\section{Basics of Geometrothermodynamics}
\label{sec:gtd}
Geometrothermodynamics (GTD) is a geometric theory of thermodynamics recently proposed by Quevedo \cite{quev07}.
It makes use of contact geometry in order to describe the phase space $\cal T$ of thermodynamic systems and express the first law of thermodynamics 
in a coordinate free fashion. Furthermore, GTD adds a Riemannian structure $G$ to the phase space and requests $G$ to be invariant
under Legendre transformations, in order to give it the same properties which one expects for ordinary thermodynamics.
Moreover, GTD introduces the manifold of the equilibrium space $\cal E$  as the maximum integral submanifold of the contact structure of $\mathcal T$,
characterized by the validity of the first law of thermodynamics \cite{quev07}. 
At the same time, GTD prescribes also to pullback the Riemannian structure $G$ to the equilibrium space. 
This results in a naturally induced Riemannian structure $g$
in $\mathcal E$, which is supposed to be the geometric counterpart of the thermodynamic system.
Such a description has been proposed in order to give thermodynamic geometry a new symmetry which was not present in previous approaches,
i.e. the Legendre invariance. 

Let us see now the mathematical definitions of the GTD objects  that we shall use in this work.
If we are given a  system with $n$ thermodynamic degrees of freedom,
we introduce first a $(2n+1)-$dimensional space ${\cal T}$ with coordinates 
$Z^A=\{\Phi, E^a, I^a\}$, with $A=0,...,2n$ and $a=1,\dots,n$, which is known as {\it the thermodynamic phase space} \cite{quev07}. 
We make use of the phase space $\mathcal T$ in order to correctly handle both the Legendre transformations and the first law of thermodynamics.
In fact, in classical thermodynamics we can change the thermodynamic potential using 
{\it a Legendre transformation}, which is defined in ${\cal T}$ as the change of coordinates given by \cite{arnold} 
\begin{equation}
\{\Phi, E^a, I^a\} \longrightarrow \{\tilde \Phi, \tilde E ^a, \tilde I ^ a\}\ ,
\end{equation}
\begin{equation}
 \Phi = \tilde \Phi - \delta_{kl} \tilde E ^k \tilde I ^l \ ,\quad
 E^i = - \tilde I ^ {i}, \ \  
E^j = \tilde E ^j,\quad   
 I^{i} = \tilde E ^ i , \ \
 I^j = \tilde I ^j \ ,
 \label{leg}
\end{equation}
where  $i\,\cup j$ can be any disjoint decomposition of the set of indices $\{1,...,n\}$,
and $k,l= 1,...,i$. 
We remark that Legendre transformations are change of coordinates in $\cal T$ and that they are not defined in the equilibrium space.
Moreover, the phase space ${\cal T}$ is equipped with a canonical contact structure called {\it the Gibbs $1$-form} defined as
\begin{equation}\label{theta1}
\Theta = d\Phi- \delta_{ab} I^b dE^a = d\Phi - I_a dE^a,
\end{equation}
which extremely resembles the first law of thermodynamics and hence will be the starting point to define the equilibrium space.

Furthermore, {\it the equilibrium space} ${\cal E}$,
is the $n$-dimensional submanifold of ${\cal T}$ defined by the embedding  
$\varphi: {\cal E} \longrightarrow {\cal T}$ under the condition 
\begin{equation}\label{defE}
\varphi^* (\Theta) =0\ , \quad {\rm i.e.} \quad d\Phi = I_a d E^a \quad {\rm and} \quad I_a = \frac{\partial \Phi}{\partial E^a}\ , 
\end{equation}
where $\varphi^*$ is the pullback of $\varphi$. 
It follows immediately from (\ref{theta1}) that (\ref{defE}) represents both the first law and the equilibrium conditions for the thermodynamic system
under analysis, so that ${\cal E}$ results to be (by definition) the submanifold of points where the first law and the equilibrium conditions hold, i.e. the geometric counterpart of the 
thermodynamic system.
It also follows that the coordinates $\{Z^A\}$ of ${\cal T}$ assume a physical meaning in ${\cal E}$. In fact, the set $\{E^a\}$, with $a=1,...,n$,
can be identified with the extensive thermodynamic variables,  while $\Phi=\Phi(E^a)$ with the fundamental equation for the thermodynamic potential
and finally the coordinates $\{I^a\}=\left\{I^a(E^a)\right\}\equiv \{\partial_{E^a}\Phi\}$, $a=1,\dots,n$, represent the intensive quantities correspoding to the extensive set $\{E^a\}$ 
(see e.g. \cite{Callen} for these definitions).

Now let us add the Riemannian structure. Since we want the Riemannian structure to share the same properties of the first law and since the first law is invariant 
under Legendre transformations, we introduce in the phase space ${\cal T}$ a metric $G$ which is invariant under Legendre transformations. In GTD there are 
several families of metrics which have this property (for a recent work on this topic see \cite{NewDev}). Among them, one has been proven particularly successful to describe
systems with second order phase transitions, as black holes are supposed to be.
Thus the candidate metric we shall use in this work is
\begin{equation}
G= \left(d\Phi - I_a dE^a\right)^2  +\Lambda
\left(\xi_{ab}E^{a}I^{b}\right)\left(\chi_{cd}dE^{c}dI^{d}\right) \  ,
\label{GII}
\end{equation}
where $\xi_{ab}$ and $\chi_{ab}$ are diagonal constant
tensors, and $\Lambda$  is an arbitrary Legendre invariant function of the coordinates $\{Z^A\}$. In particular, we choose to fix $\Lambda=1$, 
$\xi_{ab}=\delta_{ab}\equiv{\rm diag}(1,...,1)$ and  
$\chi_{ab} =\eta_{ab}\equiv {\rm diag}(-1, 1, ..., 1)$ in order to get the exact expression for the metric describing black holes phase transitions (see also \cite{PhTransGTD}).

On the other side, we are not interested in the geometric representation of the phase space, while we care about the geometric properties of the thermodynamic system, which
is paralleled by the equilibrium space $\cal E$. Thus, we pullback the metric $G$ onto $\cal E$ and obtain a Riemannian structure for the equilibrium space which 
reads
\begin{equation}\label{gII}
g^{II}\equiv \varphi^*(G)= \left(E^a \frac{\partial \Phi}{\partial E^a}\right) \left( \eta_b^{ c} \, \frac{\partial^2\Phi}{\partial E^c \partial E^d} \, d E^b d E^d \right) \ ,
\end{equation}
where $\varphi^*$ is the pullback of $\varphi$ as in (\ref{defE}) and $\eta_b^{ c}={\rm diag} (-1,1,...,1)$. 
We  remark that $g^{II}$ is (by definition) invariant under (total) Legendre transformations (see e.g. \cite{PhTransGTD}).
Moreover, we also note that $g^{II}$ can be calculated explicitly once the fundamental equation $\Phi=\Phi(E^a)$ is known.

The main thesis of GTD is that the thermodynamic properties of a system described by a fundamental equation $\Phi(E^a)$
can be translated into geometrical features of the equilibrium manifold ${\cal E}$, which in our case is described by the metric $g^{II}$. 
For example, the scalar curvature of $\cal E$ should give information about the thermodynamic interaction. This means that 
systems without interaction shall correspond to flat geometries and systems showing interaction and phase transitions should 
correspond to  curved equilibrium manifolds having curvature singularities.
It has been tested in a number of works  (see for instance \cite{quev07, PhTransGTD, hernando2,quev08,KSBH}) that indeed such correspondence works.
Furthermore, a previous work \cite{AleQuevDavood} studying the thermodynamics and GTD  of
the  Reissner-Nordstr\"om and of the Kerr black holes in any dimensions, highlighted that curvature singularities of $g^{II}$ are exactly at the same points
where the generalized susceptibilities diverge. 

In this work we extend the work in  \cite{AleQuevDavood} to the case of Myers-Perry black holes in five dimensions, with the aim of both of analyze their thermodynamic geometry 
from a new perspective
and to focus on the idea of checking what happens with a change of the potential from $\Phi=M$ to $\Phi=S$ in the GTD analysis and when the equilibrium manifold is $3$-dimensional.
The
investigation of the phase structure of Myers-Perry black holes in five dimensions is thus a matter which is interesting by itself and that will provide us with the necessary ground 
 for a new test of the
correspondence between the thermodynamic geometry $g^{II}$ of GTD and black holes thermodynamics.

\section{Myers-Perry black holes}
\label{sec:MP}

The Kerr black hole can be generalized to the case of arbitrary dimensions and arbitrary number of spins. It turns out that, provided
$d$ is the number of spacetime dimensions, then the maximum number of possible independent spins is $(d-1)/2$ if $d$ is odd and
$(d-2)/2$  if $d$ is even \cite{MPoriginal}.
Such general configurations are called Myers-Perry black holes. They deserve a special interest because they are the natural generalization
of the well-known Kerr black hole to higher number of spins and because they are shown to coexist with the Emparan-Reall black ring solution 
for some values of the parameters, thus providing the first explicit example of a violation in dimension higher than four of the uniqueness theorem 
 (see e.g. \cite{ERBR} for more details).
The line element of the Myers-Perry black hole with an arbitrary number of independent angular momenta
in Boyer-Lindquist coordinates for $d=2\,n+1$ (i.e. odd $d$) reads \cite{MPoriginal}
\begin{eqnarray}\label{RNmetric}
ds^2 &=& - d{t}^2 + {\mu r^2\over\Pi\, F}\,
 \left(d{t}  + \sum_{i=1}^n a_i\, \mu_i^2 \,d{\phi}_i \right)^2
 + {\Pi\, F\over\Pi - \mu r^2} \,dr^2
  \nonumber
+ \sum_{i=1}^n\,( r^2 + a_i^{\,2} ) \left(d\mu_i^{\,2} + \mu_i^2
\,{d{\phi}_i}^2 \right)\,,
\end{eqnarray}
with 
\begin{eqnarray}
 F &\equiv& 1-\sum_{i=1}^{n}\,{a_i^{\,2}\,\mu_i^{\,2} \over r^2 + a_i^{\,2}}\,,
  \qquad
 \Pi \equiv\, \prod_{i=1}^{n}\, (r^2 + a_i^{\,2})\,,\label{PiMP}
\end{eqnarray}
and 
\begin{eqnarray}\label{def_mu&ai}
\mu \equiv \frac{16\,\pi\,G\,M}{(d-2)\,\Omega_{(d-2)}}\,,\qquad a_i\equiv \frac{(d-2)}{2}\,\frac{J_i}{M}\,,
\end{eqnarray}

\noindent where $\Omega_{(d-2)}=\frac{2\,\pi^{n}}{\Gamma(n)}$, $M$ is the mass of the black hole, $J_i=J_1$, \dots, $J_n$ are the $(d-1)/2$ independent angular momenta and the constraint $\sum_{i=1}^n\mu_i^2=1$ holds.
Solving the equation $g^{rr}=1/g_{rr}=0$, one finds  the radius of the event horizon  (in any dimensions)
and thus derives the area and the corresponding entropy, using Bekenstein area law \cite{aman2}.

In particular, in this work we are interested in the five dimensional case, i.e. when $d=5$. 
Myers-Perry black holes in five dimensions  can have up to $2$ independent angular momenta 
and the general equation
for the area reads \cite{aman2}

\begin{equation}\label{MPArea5}
A=\frac{2\pi^2}{r_+}(r_+^2+a_1^2)(r_+^2+a_2^2)\,,
\end{equation}
where $r_+$ is the radius of the event horizon.
\noindent From the above expression the entropy can be calculated, being
\begin{equation}\label{MPS5}
S=\frac{k_B\,A}{4\,G}=\frac{1}{r_+}(r_+^2+a_1^2)(r_+^2+a_2^2)\,,
\end{equation}
where we choose $k_B$ and $G$ such that $S$ simplifies as in the second equality in (\ref{MPS5}).

Since it is rather complicated to calculate explicitly the above expression for the entropy, we will use the $M$ representation
throughout the paper. This is possible since the mass can be written in terms of $S$, $J_1$ and $J_2$ as \cite{aman2}

\begin{equation}\label{MassMP5d}
M(S,J_1,J_2)=\frac{3}{4}\,S^\frac{2}{3}\left(1+4\frac{J_1^2}{S^2}\right)^\frac{1}{3}\left( 1+4\frac{J_2^2}{S^2} \right)^\frac{1}{3}\,.
\end{equation}

\noindent Eq. (\ref{MassMP5d}) thus represents the fundamental equation for the Myers-Perry black hole in five dimensions as a thermodynamic system.
Starting from Eq. (\ref{MassMP5d}), we can analyze both the thermodynamic properties and their geometrothermodynamic counterparts.
We will split the work in order to consider the three most interesting cases, i.e. when one of the two angular momenta is zero, when they are
both non-zero but equal and finally when they are both non-zero and different among each other.

\subsection{The case $J_2=0$}

If either $J_1=0$ or $J_2=0$, we obtain the Kerr black hole in $5$ dimensions, which has been analyzed in  \cite{AleQuevDavood}. We briefly review here some of the 
results presented there and improve the analysis, including the investigation of the response functions defined in the total Legendre transformation of the mass $M$, which
we will call the Gibbsian response functions, in analogy with standard thermodynamics \cite{Callen}.
Therefore, let us suppose that $J_2=0$. 
According to our previous results \cite{AleQuevDavood}, we know that the response functions defined in the mass representation read
\begin{equation}\label{CJMP5d1}
C_{J_1}=-\frac{3S(S^2+4J_1^2)}{S^2-12J_1^2} \,,\quad
\kappa_S=\frac{3(S^2+4J_1^2)^\frac{5}{3}}{2J_1(3S^2-4J_1^2)}\,,\quad
\alpha_S=-\frac{3}{8}\frac{(S^2+4J_1^2)^\frac{5}{3}}{J_1^2\,S}\,,
\end{equation} 
where we make use of the notation $M_{E^{a}}\equiv\frac{\partial M}{\partial E^a}$ and  $M_{E^{a}E^b}\equiv\frac{\partial^2 M}{\partial E^a \partial E^b}$, for $E^i=S,J_1$.
It follows that $\alpha_S$ does not show any singularity (apart from the extremal limit $S=0$), while $C_{J_1}$
diverges at the Davies point $S^2=12J_1^2$ and $\kappa_S$ shows an additional possible phase transition
at  $3S^2=4J_1^2$. 
As it was pointed out in \cite{AleQuevDavood}, both singularities of the heat capacity and of the compressibility are in the black hole 
region and hence are physically relevant.
It was also shown that the GTD geometry (\ref{gII}) with fundamental Eq. (\ref{MassMP5d}) (with $J_2=0$) 
 is curved, indicating the presence of thermodynamic interaction,
 and that the singularities of the scalar curvature are situated  
exactly at the same points where the response functions $C_{J_1}$ and $\kappa_S$ diverge, both in the mass and in the entropy representations. 
Furthermore, it was also commented that Weinhold geometry is flat in this case and Ruppeiner thermodynamic metric diverges only in the
extremal limit $S=0$ (see e.g. \cite{aman2} for a complete analysis using these metrics).

Moreover, since the thermodynamics of black holes can depend on the chosen ensemble (see e.g. \cite{Chamb1} and \cite{Chamb2}), we now proceed to calculate  the Gibbsian response functions, which
can possibly give new information about the phase structure. 
Using the relations between thermodynamic derivatives (see \cite{Callen}), we find out that the expressions for such response functions in the 
coordinates $(S,J_1)$ used here are
\begin{equation}\label{COmMP5d1}
C_{\Omega_1}=-\frac{S(3S^2-4J_1^2)}{S^2+4J_1^2} \,,\quad
\kappa_T=-\frac{S^2-12J_1^2}{2J_1(S^2+4J_1^2)^{\frac{1}{3}}}\,,\quad
\alpha_{\Omega_1}=-\frac{8\,S}{(S^2+4J_1^2)^{\frac{1}{3}}}\,.
\end{equation} 

It is immediate to see that $C_{\Omega_1}$ never diverges and it vanishes exactly at the same points where $\kappa_S$ diverges. On the other side, $\kappa_T$ is never divergent
and it vanishes exactly where $C_{J_1}$ diverges, while $\alpha_{\Omega_1}$ is always finite. It follows that the Gibbsian response functions do not add any information to the knowledge 
of the phase structure of this configuration, as they change sign exactly at the points that we have already analyzed,
 therefore we conclude that the divergences of the scalar curvature of the metric (\ref{gII}) match exactly the points of second order phase transitions. 

Let us now add a second spin parameter and show that there is still a concrete correlation
between the geometric description performed with $g^{II}$ and the thermodynamic properties.
To do so, we first focus on the special case of equation (\ref{MassMP5d}) in which $J_1=J_2=J$,
 and afterwards we will consider the completely general case, 
i.e. with $J_1$ and
$J_2$ both different from zero and from each other. 
In particular, in the latter case we will get a $3$-dimensional thermodynamic manifold labelled by $(E^1=S, E^2=J_1,E^3=J_2)$ and hence we will consider the $3$-dimensional version of 
the metric (\ref{gII}).

\subsection{The case $J_1=J_2\equiv J$}
Another  special case in Eq. (\ref{MassMP5d}) which is of interest is the case in which the two angular momenta are fixed to be equal, i.e. $J_1=J_2\equiv J$.
 This is interesting
from the mathematical and physical point of view since it is the only case in which the angular momenta are both different  from zero and at the same time
 the thermodynamic
manifold is $2$-dimensional.
In fact, the mass fundamental equation (\ref{MassMP5d}) in this case is given by

\begin{equation}\label{MassMP5d2equal}
M(S,J)=\frac{3}{4}\,S^\frac{2}{3}\left(1+4\frac{J^2}{S^2}\right)^\frac{2}{3}\,,
\end{equation}

\noindent and the response functions can then be accordingly calculated, to give
\begin{equation}\label{CJMP5d2equal}
C_{J}=-\frac{3S(S^4-16J^4)}{S^4-32J^2S^2-80J^4} \,, \quad
\kappa_S=\frac{3S^\frac{2}{3}(S^2+4J^2)^\frac{4}{3}}{4J(3S^2+4J^2)}\,,\quad
\alpha_S=-\frac{3}{16}\frac{S^\frac{5}{3}(S^2+4J^2)^\frac{4}{3}}{J^2(S^2+2J^2)}\,.
\end{equation} 

\noindent From (\ref{CJMP5d2equal}), it follows that in this case $\alpha_S$ and $\kappa_S$  do not show any singularity, while $C_{J}$
diverges at the roots of the denominator $\mathcal D_C=S^4-32J^2S^2-80J^4$.
We also observe that the temperature of this black hole is given by 
\begin{equation}\label{entropycase2}
T\equiv\left(\frac{\partial M}{\partial S}\right)_J=\frac{1}{2}\frac{S^2-4J^2}{S^{5/3}(S^2+4J^2)^{1/3}}\,,
\end{equation} 
therefore the extremal limit $T=0$ is reached when $\frac{J^2}{S^2}=\frac{1}{4}$.

Solving the Eq. $\mathcal{D}_C= 0$, we find that the singularities of the heat capacity are situated at a value $S_{\rm critical}$ for the entropy such that 
\begin{equation}\label{Scritical}
\left.\frac{J^2}{S^2}\right|_{S=S_{\rm critical}}=\frac{\sqrt{21}-4}{20}\,,
\end{equation} 
which is less
than the extremal limit. Therefore Davies point of phase transition belongs to the black hole region and we shall investigate it.

It is convenient also in this case to write the full set of thermodynamic response functions, including the Gibbsian ones. Again, making use of the 
relations between thermodynamic derivatives, we find out that they read
\begin{equation}\label{COmMP5d2equal}
C_{\Omega}=-\frac{S\,(S^2-4J^2)(3S^2+4J^2)}{(S^2+4J^2)^2} \,,\quad
\kappa_T=-\frac{S^{2/3}}{4\,J}\frac{\mathcal D_C}{(S^2+4J^2)^{5/3}} \,,\quad
\alpha_{\Omega}=\frac{8\,S^{5/3}\,(S^2+2J^2)}{(S^2+4J^2)^{5/3}} \,.
\end{equation} 
In this case we observe that the only divergence of the response functions in (\ref{CJMP5d2equal}), i.e. the divergence of $C_J$, is again controlled by the vanishing
of $\kappa_T$. Furthermore, both $C_J$ and $C_\Omega$ vanish at the extremal limit, but this does not correspond to any divergence of $\kappa_S$, hence we expect
the curvature of the thermodynamic metric to diverge only at the points where $\mathcal D_C=0$.

From the point of view of Geometrothermodynamics, given the fundamental equation (\ref{MassMP5d2equal})  and the general metric (\ref{gII}),
we can calculate the particular metric and the scalar curvature for the equilibrium manifold of the MP black hole with two equal angular momenta, both in the mass and
in the entropy representations.

\noindent The metric in the $M$ representation reads
\begin{equation}
g^{II}_M=\frac{1}{S^{4/3}(S^2+4J^2)^{2/3}}\bigg\{\frac{1}{12}\frac{\cal D_C}{S^2} \, dS^2+\frac{2 (3S^2+4J^2)}{3}\, dJ^2\bigg\}\,,
\end{equation}
therefore its scalar curvature is
\begin{equation}
R_M=\frac{24 S^{10/3} (S^2+4J^2)^{2/3}(5\,S^6+48\,J^2\,S^4-368\,J^4\,S^2-896\,J^6)}{\mathcal D_C^2 (3S^2+4J^2)^2}\,.
\end{equation}
The numerator is a not very illuminating function that never vanishes when the denominator is zero and $\mathcal D_C$ is exactly the denominator of the heat capacity $C_J$. 
Therefore, the singularities of $R_M$ correspond exactly to those of  $C_J$ (resp. to the vanishing of $\kappa_T$). 
We remark that the factor $3S^2+4J^2$ in the denominator of $R_M$, though being always 
different from $0$ (thus not indicating any phase transition in this case), is exactly the denominator of the compressibility $\kappa_S$ (resp. a factor in the numerator of $C_\Omega$). 

To continue with the analysis, in \cite{NewDev} it was presented a general relation (see Eq. (36) therein) to express $g^{II}$ with $\Phi=S$ (i.e. in the $S$ representation) in the coordinates of the $M$ representation
(i.e. $\{E^a\}=(S,J)$). Such relation in the present case reads
\begin{equation}\label{conformalrel}
g^{II}_S=\frac{M-J\,\Omega}{T^3} \left[T\, M_{SS} dS^2+2\,\Omega\,M_{SS}dS\,dJ+(2\,\Omega\,M_{SJ}-T\,M_{JJ})\,dJ^2\right]\,,
\end{equation}
where $T\equiv\frac{\partial M}{\partial S}$ is the temperature, $\Omega\equiv\frac{\partial M}{\partial J}$ is the angular velocity at the horizon and $M_{E^aE^b}\equiv\frac{\partial^2 M}{\partial E^a \partial E^b}$, for $E^i=S,J$. 
Using Eq. (\ref{conformalrel}) and Eq. (\ref{MassMP5d2equal}) for the mass in terms of $S$ and $J$,
we can calculate the expression for metric $g^{II}_S$ in the coordinates $(S,J)$, which reads
\begin{eqnarray}\label{gScaseII}
g^{II}_S&=&\frac{1}{3 (S^2+4J^2)(S+2J)^2(S-2J)^2}
\bigg\{ -\frac{3S^2-4J^2}{2} \mathcal{D}_C\, dS^2+\\
&+&\frac{8SJ (3S^2-4J^2)}{(S+2J)(S-2J)} \mathcal{D}_C\,  dS\,dJ
-4S^2\frac{9S^6+156S^4J^2+112S^2J^4-448J^6}{(S+2J)(S-2J)} \, dJ^2
\bigg\}\,.\nonumber
\end{eqnarray}
Consequently, the scalar curvature is
\begin{equation}\label{R_SequalJ}
R_S=\frac{\mathcal N_S}{(3S^2-4J^2)^3(S^2+4J^2)^2\,\mathcal{D}_C^2}\,,
\end{equation}
where $\cal N_S$ is again a function which never vanishes at the points where the denominator is zero.
 From Eq. (\ref{R_SequalJ}), we see that the denominator of $C_J$ is present in the denominator of $R_S$. 
Furthermore, the factor $S^2+4J^2$ is never zero, hence it does not give any additional singularity.
On the other hand, the factor $3S^2-4J^2$ is clearly vanishing when $\frac{J^2}{S^2}=\frac{3}{4}$, which is readily greater than the extremal limit 
$\frac{J^2}{S^2}=\frac{1}{4}$ and hence it has no physical relevance in our analysis.

We thus conclude that also in this case the GTD geometry $g^{II}$ exactly reproduces the phase transition structure
of the Myers-Perry black holes both in the mass and in the entropy representation. We comment that in the entropy representation
there is an additional singularity which does not correspond to any singularity of the response functions.  
However, such singularity is situated out of the black hole region and thus it is not to be considered here.
We also remark that Ruppeiner curvature in this case reads 
$R=-\frac{S\,(S^2+12J^2)}{S^4-16J^4}$ and hence it diverges 
only in the extremal limit, while Weinhold metric is flat.

In the next subsection we will analyze the general case of the Myers-Perry black hole in five dimensions, i.e. when the two angular momenta are
allowed to vary freely.

\subsection{The general case in which $J_1\neq J_2\neq0$}
Perhaps the most interesting  case is the most general one, in which the two angular momenta are allowed to vary freely. In this 
case the thermodynamic manifold is $3$-dimensional
and the mass fundamental equation is given by (\ref{MassMP5d}).

The generalized susceptibilities can then be accordingly calculated.
The heat capacity at constant angular momenta $J_1$ and $J_2$ reads 
\begin{equation}\label{CJMP5d2different}
C_{J_1,J_2}=\frac{M_S}{M_{SS}}=-\frac{3\,S\,(S^2+4J_1^2)(S^2+4J_2^2)(S^4-16J_1^2J_2^2)}
{\mathcal D_C} \,,
\end{equation} 
where 
\begin{equation}\label{D_C}
\mathcal D_C=S^8-12(J_1^2+J_2^2)S^6-320J_1^2J_2^2S^4-576J_1^2J_2^2(J_1^2+J_2^2)S^2-1280J_1^4J_2^4\,.
\end{equation}
Furthermore, one can define the $3$ analogues of the adiabatic compressibility
 as
\begin{equation}\label{KSMP5d2different}
(\kappa_S)_{11}\equiv\left(\frac{\partial J_1}{\partial \Omega_1}\right)_S =\frac{3S^\frac{2}{3}(S^2+4J_1^2)^\frac{5}{3}}{2(S^2+4J_2^2)^\frac{1}{3}(3S^2-4J_1^2)}\,,\quad
(\kappa_S)_{22}\equiv\left(\frac{\partial J_2}{\partial \Omega_2}\right)_S =\frac{3S^\frac{2}{3}(S^2+4J_2^2)^\frac{5}{3}}{2(S^2+4J_1^2)^\frac{1}{3}(3S^2-4J_2^2)}\,,
\end{equation} 
and
\begin{equation}\label{alpha_SMP5different}
(\kappa_S)_{12} \equiv\left(\frac{\partial J_1}{\partial \Omega_2}\right)_S =\frac{3}{16}\,{\frac {{S}^{2/3}{ \left( {S}^{2}+4\,J_1^{2} \right) }^{2
/3}{ \left( {S}^{2}+4\,J_2^{2} \right) }^{2/3}}{{ J_1}\,{J_2}}}\,.
\end{equation} 
\noindent Finally, the analogues of the expansion are given by
\begin{equation}\label{alpha_SMP5d2different}
\alpha_{S,J_2}\equiv\left(\frac{\partial J_1}{\partial T}\right)_S =-\frac{3}{8}\frac{S^\frac{5}{3}(S^2+4J_1^2)^\frac{5}{3}(S^2+4J_2^2)^\frac{2}{3}}{J_1(S^4+6S^2J_2^2+8J_1^2J_2^2)}\,,
\quad
\alpha_{S,J_1}\equiv\left(\frac{\partial J_1}{\partial T}\right)_S =-\frac{3}{8}\frac{S^\frac{5}{3}(S^2+4J_2^2)^\frac{5}{3}(S^2+4J_1^2)^\frac{2}{3}}{J_2(S^4+6S^2J_1^2+8J_1^2J_2^2)}\,.
\end{equation} 

In this case neither $(\kappa_S)_{12}$ nor the expansions show any singularity, while $C_{J_1,J_2}$
diverges when $\mathcal D_C=0$ and the compressibilities $(\kappa_S)_{11}$ and $(\kappa_S)_{22}$
 diverge when $3S^2-4J_1^2=0$ and $3S^2-4J_2^2=0$ respectively.
Furthermore, the temperature reads
\begin{equation}\label{TMP2J}
T=\frac{1}{2S^{5/3}}\frac{S^4-16J_1^2J_2^2}{(S^2+4J_1^2)^{2/3}(S^2+4J_2^2)^{2/3}}\,,
\end{equation}
hence the extremal limit is reached for $\frac{J_1J_2}{S^2}=\frac{1}{4}$. 
The heat capacity diverges when $\mathcal{D}_C = 0$, which is an algebraic equation of degree $8$ in $S$. 
We can solve numerically such equation and obtain 
the critical value $S={\rm Scritical}$ in terms of $J_1$ and $J_2$. Taking only  
the roots which are real and positive, we can compare them with the extremal limit by doing
\begin{equation}\label{confronto1}
\left.\frac{J_1J_2}{S^2}\right|_{S={\rm Sextremal}}-\left.\frac{J_1J_2}{S^2}\right|_{S={\rm Scritical}}=\frac{1}{4}-\left.\frac{J_1J_2}{S^2}\right|_{S={\rm Scritical}}\,.
\end{equation}
The plot of the result is given in Fig. 1 for some values of $J_1$ and $J_2$.
\begin{figure}\label{confronto1fig}
\begin{center}
\includegraphics[width=3.5in]{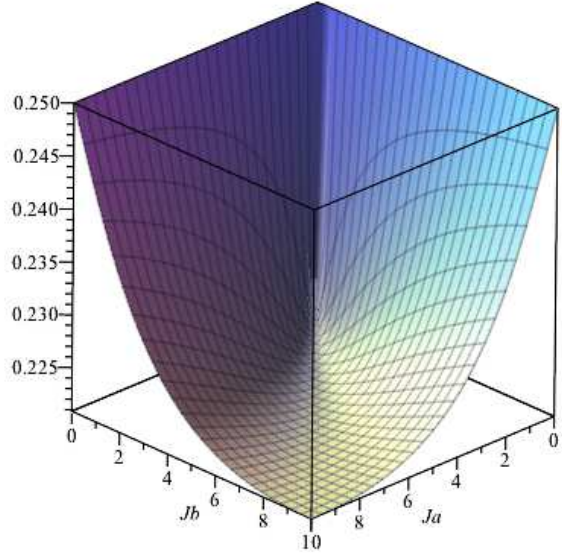}
\caption{The difference between the extremal limit ($\frac{J_1J_2}{S^2}=\frac{1}{4}$) and the value of $\frac{J_1J_2}{S^2}$ at the critical point of the heat capacity,
 plotted for values of $J_1$ and $J_2$  in the interval $[0,10]$.}
\end{center}
\end{figure}
As we can see from Fig. 1, the difference in Eq. (\ref{confronto1}) is always positive, hence the point of phase transition signaled by the divergence of the heat capacity 
is always in the black hole region.

In the same way, we can solve $3S^2-4J_1^2=0$ and see whether the divergence of $(\kappa_S)_{11}$ lies in the black hole region or not.
It turns out that the denominator of $(\kappa_S)_{11}$ vanishes for values of $S$ such that  $\frac{J_1^2}{S^2}=\frac{3}{4}$, 
which means that $\frac{J_1J_2}{S^2}=\frac{3}{4}\frac{J_2}{J_1}$.
Therefore, we have that $\frac{1}{4}-\frac{3}{4}\frac{J_2}{J_1}$ is positive provided $J_1>3J_2$ for $J_1>0$ or $J_1<3J_2$ for $J_1<0$. 
Summing up, the divergences of $(\kappa_S)_{11}$
can be in the black hole region for appropriate values of $J_1$ and $J_2$. Analogously, the divergences of $(\kappa_S)_{22}$ can also be in the black hole region.

As in the preceding sections, we will now focus on the Gibbsian response functions, in order to make the analysis complete. The heat capacity at 
constant angular velocities read
\begin{equation}\label{COmMP5d2different}
C_{\Omega_1,\Omega_2} \equiv T \left(\frac{\partial S}{\partial T}\right)_{\Omega_1,\Omega_2}= \frac{-3\,S(S^4-16J_1^2J_2^2)(3S^2-4J_1^2)(3S^2-4J_2^2)(S^2+4J_1^2)(S^2+4J_2^2)}{\mathcal D(S,J_1,J_2)}\,,
\end{equation} 
where the denominator is given by
\begin{equation}
\begin{split}
\mathcal D(S,J_1,J_2) & = \, 9S^{12}+72(J_1^2+J_2^2)S^{10}+16(9J_1^4+95J_1^2J_2^2+9J_2^4)S^8+5376J_1^2J_2^2(J_1^2+J_2^2)S^6+\\
&-256J_1^2J_2^2(9J_1^4-101J_1^2J_2^2+9J_2^4)S^4-6144J_1^4J_2^4(J_1^2+J_2^2)S^2-53248J_1^6J_2^6\,.
\end{split}
\end{equation}
Furthermore, one can define three generalized susceptibilities, analogous to the isothermal compressibility, as
\begin{equation}\label{KTMP5d2different1}
(\kappa_T)_{11} \equiv \left(\frac{\partial J_1}{\partial \Omega_1}\right)_T\,, \qquad 
(\kappa_T)_{12} \equiv \left(\frac{\partial J_1}{\partial \Omega_2}\right)_T\,, \qquad
(\kappa_T)_{22} \equiv \left(\frac{\partial J_2}{\partial \Omega_2}\right)_T\,.
\end{equation} 
\noindent For the Myers-Perry black hole they can be written as
\begin{eqnarray}\label{KTMP5d2different2}
(\kappa_T)_{11}  &=& -\frac{S^{2/3}}{2}\frac{\mathcal D_C}{(S^2+4J_1^2)^{1/3}(S^2+4J_2^2)^{1/3}(S^6-12J_2^2S^4+48J_1^2J_2^2S^2+192J_1^2J_2^4)}\,,\\
(\kappa_T)_{22} &=& -\frac{S^{2/3}}{2}\frac{\mathcal D_C}{(S^2+4J_1^2)^{1/3}(S^2+4J_2^2)^{1/3}(S^6-12J_1^2S^4+48J_1^2J_2^2S^2+192J_2^2J_1^4)}\,,
\end{eqnarray} 
while $(\kappa_T)_{12}$ has a more cumbersome expression and we will not write it here, since it has the same properties of $(\kappa_T)_{11}$ and $(\kappa_T)_{22}$
for as regards to our analysis, i.e. it is proportional to the denominator 
of $C_{J_1,J_2}$ defined in (\ref{D_C}) and it has a non-trivial denominator
(one can also introduce the two analogues of the thermal expansion, but for the sake of simplicity we are not going to write them here, since they do not show any singularities and hence they do not
play any role in our analysis). 

Therefore, from the thermodynamic point of view, we remark that the divergences of $C_{J_1,J_2}$ are matched by the vanishing of the three quantities $(\kappa_T)_{ij}$, while
the divergences of $(\kappa_S)_{11}$ and $(\kappa_S)_{22}$ are reproduced as zeroes of the heat capacity $C_{\Omega_1,\Omega_2}$. This behavior is in agreement with 
the analysis of the preceding sections.
Furthermore, in this case the heat capacity $C_{\Omega_1,\Omega_2}$ and the generalized compressibilities $(\kappa_T)_{ij}$ possibly show additional phase transitions,
which is a further indication of the fact that black holes exhibit  different thermodynamic behavior in different potentials.

Now let us turn to the GTD analysis. Given the fundamental equation  (\ref{MassMP5d})  and the general metric (\ref{gII}),
we can calculate the particular metric and the scalar curvature for the MP black hole with two free angular momenta, both in the mass and
in the entropy representations.
\noindent The metric in the $M$ representation reads
\begin{eqnarray}\label{gIIM3D}
g^{II}_M&=&\frac{1}{3 S^{4/3} (S^2+4J_1^2)^{1/3} (S^2+4J_2^2)^{1/3}}
\bigg\{
-\frac{1}{4}\frac{\cal D_C}{S^2 (S^2+4J_1^2) (S^2+4J_2^2)}\,dS^2+\\
&+&\frac{(3S^2-4J_1^2)(S^2+4J_2^2)}{S^2+4J_1^2}\,dJ_1^2 +\frac{(3S^2-4J_2^2)(S^2+4J_1^2)}{S^2+4J_2^2}\,dJ_2^2+16J_1J_2\,dJ_1\,dJ_2
\bigg\}\,,\nonumber
\end{eqnarray}
hence its scalar curvature is
\begin{equation}
R_M=\frac{\mathcal N_M}{\mathcal D_C^2 \left[3S^4-4S^2(J_1^2+J_2^2)-16J_1^2J_2^2\right]^2(S^2+4J_1^2)^{2/3}(S^2+4J_2^2)^{2/3}}\,,
\end{equation}
where $\mathcal D_C$ is as usual the denominator of $C_{J_1,J_2}$ defined in Eq. (\ref{D_C}). Since there is no term in the numerator $\mathcal N_M$ which cancels out
the divergences that happen when $\mathcal D_C=0$, we can conclude that every phase transition related to the heat capacity $C_{J_1,J_2}$ is properly reproduced by the scalar 
curvature $R_M$.
In addition, in this case the factor $3S^4-4S^2(J_1^2+J_2^2)-16J_1^2J_2^2$ can also vanish, 
possibly giving an additional singularity which does not correspond to the ones shown by the response functions. 
It is easy to calculate that  $3S^4-4S^2(J_1^2+J_2^2)-16J_1^2J_2^2=0$ for values of $S$ such that
 \begin{equation}\label{possiblenewsing}
 \frac{J_1J_2}{S^2}=\frac{1}{8}\frac{\left(-J_1^2-J_2^2+\sqrt{J_1^4+14J_1^2J_2^2+J_2^4}\right)}{J_1J_2}\,.
 \end{equation}
 We can thus calculate the difference between the extremal limit $\frac{J_1J_2}{S^2}=\frac{1}{4}$ and the  critical value (\ref{possiblenewsing}). The result is 
 \begin{eqnarray}\label{difference}
\frac{1}{4}&-&\frac{1}{8}\frac{\left(-J_1^2-J_2^2+\sqrt{J_1^4+14J_1^2J_2^2+J_2^4}\right)}{J_1J_2}=\\
&=&-\frac{1}{48}\frac{\left(J_1^2+J_2^2-6J_1J_2+\sqrt{J_1^4+14J_1^2J_2^2+J_2^4}\right)\left(J_1^2+J_2^2-\sqrt{J_1^4+14J_1^2J_2^2+J_2^4}\right)}{J_1^2J_2^2}\nonumber\,,
 \end{eqnarray}
 which can be positive for appropriate values of $J_1$ and $J_2$. Therefore such points of divergence of $R_M$ are in the black hole region for some values of the parameters.
Hence we conclude  that the behavior of $R_M$ perfectly matches
the behavior of $C_{J_1,J_2}$, but in this case it does not reproduce the additional possible phase transitions indicated by the singularities of the compressibilities $(\kappa_S)_{11}$
and $(\kappa_S)_{22}$
and possibly
shows some additional unexpected singularities.

However, we can give  a precise physical meaning to such additional singularities. In fact, if we evaluate the determinant of the Hessian of the mass
with respect to the angular momenta $J_1$ and $J_2$, we get
\begin{equation}\label{HessianM}
{\rm det}( \,{\rm Hess}(M)_{J_1J_2})\equiv M_{J_1J_1}M_{J_2J_2}-M_{J_1J_2}^2=\frac{4}{3}\frac{3S^4-4S^2(J_1^2+J_2^2)-16J_1^2J_2^2}{S^{4/3}(S^2+4J_1^2)^{4/3}(S^2+4J_2^2)^{4/3}}\,,
\end{equation}
from which we can see that the numerator is exactly the factor in the denominator of $R_M$ whose roots give the additional singularities.
Since the Hessians of the energy in thermodynamics are related to the stability conditions, we suggest that the physical meaning of such additional divergences of $R_M$
is to be found in a change of stability of the system, e.g. from a stable phase to an unstable one.

On the other side, using the relation (\ref{conformalrel}) for $g^{II}$ between the $M$ and the $S$ representation, naturally extended to the $3$-dimensional case with coordinates $(S,J_1,J_2)$, i.e.
\begin{eqnarray}\label{conformalrel3D}
g^{II}_S&=&\frac{M-J_1\,\Omega_1-J_2\Omega_2}{T^3} \bigg[T\, M_{SS} dS^2+2\,\Omega_1\,M_{SS}dS\,dJ_1+\\
&+&2\,\Omega_2\,M_{SS}dS\,dJ_2+(2\,\Omega_1\,M_{SJ_1}-T\,M_{J_1J_1})\,dJ_1^2
+(2\,\Omega_2\,M_{SJ_2}-T\,M_{J_2J_2})\,dJ_2^2+\nonumber\\
&-2&(T\,M_{J_1J_2}-\Omega_1\,M_{SJ_2}-\Omega_2\,M_{SJ_1})\,dJ_1\,dJ_2\nonumber
\bigg]\,,
\end{eqnarray}
we can now calculate the metric in the $S$ representation, which reads
\begin{eqnarray}\label{gIISMP3D}
g^{II}_S&=&\frac{\left[3S^4+4S^2(J_1^2+J_2^2)-16J_1^2J_2^2\right] \mathcal D_C}{3(S^2+4J_1^2)(S^2+4J_2^2)(S^2-4J_1J_2)^2(S^2+4J_1J_2)^2}
\bigg\{
\frac{1}{2}\,dS^2+\\
&+&4\frac{SJ_1(S^2+4J_2^2)}{(S^2-4J_1J_2)(S^2+4J_1J_2)}\,dS\,dJ_1+4\frac{SJ_2(S^2+4J_1^2)}{(S^2-4J_1J_2)(S^2+4J_1J_2)}\,dS\,dJ_2+\nonumber\\
&-&2\frac{S^2(S^2+4J_2^2)^2\left[3S^6+26S^4J_1^2+144S^2J_1^2J_2^2+320J_1^4J_2^2\right]}{\mathcal D_C(S^2-4J_1J_2)(S^2+4J_1J_2)}\,dJ_1^2+\nonumber\\
&-&2\frac{S^2(S^2+4J_1^2)^2\left[3S^6+26S^4J_2^2+144S^2J_1^2J_2^2+320J_2^4J_1^2\right]}{\mathcal D_C (S^2-4J_1J_2)(S^2+4J_1J_2)}\,dJ_2^2+\nonumber\\
&-&32\frac{S^2J_1J_2(S^2+4J_1^2)(S^2+4J_2^2)\left[5S^4+12S^2(J_1^2+J_2^2)+16J_1^2J_2^2\right]}{\mathcal D_C (S^2-4J_1J_2)(S^2+4J_1J_2)}\,dJ_1\,dJ_2\nonumber
\bigg\}\,.
\end{eqnarray}
The scalar curvature can thus be calculated to obtain
\begin{equation}
R_S=\frac{\mathcal N_S}{\mathcal D_C^2 \left[3S^4+4S^2(J_1^2+J_2^2)-16J_1^2J_2^2\right]^3S^2(S^2+4J_1^2)(S^2+4J_2^2)}\,.
\end{equation}

\noindent In this case we see again that the denominator of the heat capacity $\mathcal D_C$ is present in the denominator of $R_S$. 
Furthermore, the second factor, which is slightly different from the factor in the denominator of $R_M$,  vanishes 
 for values of $S$ such that
 \begin{equation}\label{possiblesing2}
 \frac{J_1J_2}{S^2}=\frac{1}{8}\frac{J_1^2+J_2^2+\sqrt{J_1^4+14J_1^2J_2^2+J_2^4}}{J_1J_2}\,.
 \end{equation} 
 The above discussion for the additional singularity of $R_M$ does not apply in this case, since one can easily show that the points
 described by (\ref{possiblesing2})
 do not belong to the black hole region for any values of $J_1$ and $J_2$. 
 However, we comment in passing that such additional singularities are still related to the vanishing of the determinant of the
 Hessian of the entropy $S$ with respect to the angular momenta $J_1$ and $J_2$. 
 Therefore they still indicate the points where the Hessian vanishes, although
 they are not situated in the black hole region in this case. 
We infer from these results that the physical meaning of the divergences of the scalar curvature of the metric $g^{II}$ for  such a $3$-dimensional equilibrium manifold
is related to the divergences of the heat capacity at constant angular momenta and to the zeroes of the Hessian of the potential with respect to those momenta, both
in the mass and in the entropy representation. On the other side, from the full analysis of the divergences of the generalized response functions, we see that there are other
possible points of phase transitions, related to divergences of the compressibilities, which appear to be not enclosed by the analysis given with $g^{II}$. We also comment 
that we could have used the potential $\Phi=G\equiv M-T\,S-J_1\,\Omega_1-J_2\,\Omega_2$ 
in writing the metric (\ref{gII}) to study the GTD analysis in the $G$ representation, but such investigation would have lead
to exactly the same results, as it has to be, since the metric (\ref{gII}) is invariant under total Legendre transformations.
 
To conclude, we observe that in \cite{aman2} the case of the full Myers-Perry black hole thermodynamics has been investigated using Weinhold and Ruppeiner
thermodynamic geometries. The authors proved that both Weinhold and Ruppeiner scalar curvatures only diverge in the extremal limit.


\section{Conclusions}
\label{sec:con}

In this work we have analyzed the thermodynamics and thermodynamic geometry   of different Myers-Perry black holes configurations in five dimensions, classifying them
according to the values of the two possible independent angular momenta.

To this end, we followed the approach of Davies for the standard analysis of the thermodynamic properties in different potentials and used the approach of GTD for the thermodynamic geometric investigation.
The present work has been carried out with the twofold aim of understanding the phase structure of Myers-Perry black holes in five dimensions
and  infer new conclusions on the physical meaning of the metric $g^{II}$, both in the mass and in the entropy representations.

Our results indicate that the Myers-Perry black holes in five dimensions have a non-trivial phase structure in the sense of Davies. 
In particular, the analysis of the response functions indicate that both the heat capacities and the compressibilities defined in the $M$ potential diverge at some points,
which is usually interpreted as the hallmark of a phase transition. Interestingly, such a behavior is matched by the vanishing of the corresponding Gibbsian response functions
in all the cases studied here.
Moreover, in the most general case when the two angular momenta vary freely, we have shown that the Gibbsian response functions provide some additional singularities, indicating
that the analysis in the $M$ potential is different from that performed in the $G$ potential.

In all the cases studied in this work, the phase transitions 
are well reproduced by the GTD analysis, while they 
are not reproduced by the thermodynamic geometries of Weinhold and Ruppeiner, whose analysis has been observed to correspond to other approaches (see e.g. \cite{ArcioniTalla}).
We have also  found that the scalar curvature of the metric $g^{II}$ shows a very similar behavior in the $M$ representation to that of the $S$ 
representation. In particular, for the cases in which we have only two degrees of freedom we argue that no physical difference has been detected and we have shown that
not only the phase transitions indicated by $C_J$ are reproduced, but also the ones indicated by divergences of $\kappa_S$. 
Moreover, a detailed analysis of the Gibbsian response functions showed that such divergences correspond to points where $\kappa_T$ and $C_\Omega$ vanish
and change their character. We therefore conclude that for such cases the divergences of the scalar curvature of $g^{II}$ reproduce the full set of second order phase transitions
considered here.

On the other side, it seems that analyzing the general case in which both angular momenta are switched on, i.e. a thermodynamic system with 
three degrees of freedom, some differences might appear.  
In fact, the phase transitions signaled by $C_{J_1,J_2}$ are still obtained as curvature singularities in both representations.
Nevertheless, the scalar curvature has some additional divergences, which for the case of the $M$ representation can be in the black hole region for appropriate values of the angular momenta and that
apparently are not directly related to the response functions of the system.
However, we claim that such additional divergences are linked to the vanishing of the Hessian determinant of the potential $M$ with respect to the two angular momenta, 
therefore they mark the transition from a stable phase to an unstable one. In our opinion this means that the physical meaning of the scalar curvature of the metric $g^{II}$ for thermodynamic
systems with three degrees of freedom goes beyond the well established correspondence with the generalized susceptibilities, i.e. second order derivatives of the potential, 
encompassing also questions of stability related to their mutual relation, i.e. determinants of the Hessian of the potential.
This is also supported by the analysis of the scalar curvature in the $S$ representation, which again shows singularities exactly at those points where the Hessian of the entropy with respect to the 
two angular momenta vanishes, so from the mathematical point of view the situation is basically the same. It is interesting however to note that in the $S$ representation such points are not in the black
hole  region, a direct evidence of the fact that black hole thermodynamics strictly depends on the potential being used. 
Moreover, in the completely general case, some additional divergences appear when considering the Gibbsian response functions, which are not present in the thermodynamic analysis in the $M$ potential, 
nor are
indicated as curvature singularities of $g^{II}$. The study of such additional singularities goes beyond the scope of this work and may be the matter of further investigation.
We also expect to extend this work in the nearest future and find a number of further examples which support (or discard) the interpretation of the thermodynamic metric $g^{II}$ for thermodynamic systems with
$3$ degrees of freedom given here.  

\section*{Acknowledgements}
The authors want to thank prof. H. Quevedo for insightful suggestions. A.B. wants to thank ICRA for financial support. 


\begin{thebibliography}{99}

\bibitem{Hawking} S. W. Hawking,  Commun. Math. Phys. {\bf 43}, 199, (1975).

\bibitem{Bekenstein} J. D. Bekenstein, Phys. Rev. D {\bf 7}, 2333, (1973).

\bibitem{Bardeen} J. M. Bardeen, B. Carter, and S.W. Hawking,  Commun. Math. Phys. {\bf 31}, 161, (1973).

\bibitem{ArcioniTalla} G. Arcioni and E. Lozano-Tellechea, Phys. Rev. D {\bf 72}, 104021, (2005). 

\bibitem{Davies}
P. C. W. Davies, Rep. Prog. Phys. \textbf{41}, 1313, (1978).

\bibitem{Lousto}
C. O. Lousto, Nucl. Phys. B {\bf 410}, 155-172,  (1993); Erratum-ibid. B {\bf 449}, 433,  (1995).


\bibitem{Lousto2}
C. O. Lousto, Int. J. Mod. Phys. D {\bf 6}, 575-590, (1997).

\bibitem{cai1} R. G. Cai and J. H. Cho, Phys. Rev. D \textbf{60}, 067502 (1999).


\bibitem{cai2} J. Shen, R. G. Cai, B. Wang and R. K. Su, Int. J. Mod. Phys. A {\bf 22}, 11-27, (2007).


\bibitem{Bana1}
R. Banerjee, S. K. Modak and S. Samanta, Eur. Phys. J. C {\bf 70}, 317-328, (2010).

\bibitem{Bana1bis}
R. Banerjee, S. K. Modak and S. Samanta, Phys. Rev. D {\bf 84}, 064024, (2011).

\bibitem{Bana1ter}
R. Banerjee and D. Roychowdhury, JHEP 1111:004, (2011).

\bibitem{Bana1quater}
R. Banerjee, S. K. Modak and D. Roychowdhury, JHEP 1210:125, (2012).


\bibitem{Bana2}
R. Banerjee, S.  Ghosh and D. Roychowdhury, Phys. Lett. B {\bf 696}, 156, (2011).

\bibitem{Bana2bis}
R. Banerjee and D. Roychowdhury, Phys. Rev. D {\bf 85}, 104043, (2012).

\bibitem{Bana2ter}
R. Banerjee and D. Roychowdhury, Phys. Rev. D {\bf 85}, 044040, (2012).

\bibitem{Bana2quater}
A. Lala and D. Roychowdhury, Phys. Rev. D {\bf 86}, 084027 (2012).



\bibitem{Gibbs} J. W. Gibbs, {\it The collected works}, Vol. 1, Thermodynamics (Yale 
University Press, 1948).


\bibitem{cara} C. Caratheodory, {\it Untersuchungen \"uber die Grundlagen der 
Thermodynamik},   Math. Ann. {\bf 67}, 355, (1909).  

\bibitem{fisher} R. A. Fisher,  Proc. Camb. Phil. Soc. {\bf 22}, 700, (1925).

\bibitem{rao} C. R. Rao, Bull. Calcutta Math. Soc. {\bf 37}, 81, (1945).

\bibitem{brody} D. C. Brody and D. W. Hook, J. Phys. A: Math. Theor. {\bf 42}, 023001, (2009).

\bibitem{wein} F. Weinhold, {\it Metric geometry of equilibrium thermodynamics I, II, III, IV}, J. Chem. Phys. {\bf 63}, 2479, 2484, 2488, 2496, (1975).

\bibitem{rupp1} G. Ruppeiner, Phys. Rev. A {\bf 20}, 1608, (1979).

\bibitem{rupp2} G. Ruppeiner,  Rev. Mod. Phys. {\bf 67}, 605, (1995).

\bibitem{rupp3} G. Ruppeiner, Phys. Rev. D {\bf 78}, 024016, (2008);

\bibitem{aman1} J. E. \AA man, I. Bengtsson and Narit Pidokrajt, Gen. Rel. Grav. {\bf 38}, 1305-1315, (2006).

\bibitem{aman2}
J. E. \AA man and N. Pidokrajt, Phys. Rev. D {\bf 73}, 024017, (2006).


\bibitem{laszlo}
L. A. Gergely, N. Pidokrajt and S. Winitzki, Eur. Phys. J. C {\bf 71}, 1569, (2011).

\bibitem{quev07} H. Quevedo, J. Math. Phys. {\bf 48}, 013506, (2007).

\bibitem{hernando2} H. Quevedo, Gen. Rel. Grav. {\bf 40}, 971, (2008).


\bibitem{PhTransGTD} 
H. Quevedo, A. S\'anchez, S. Taj, and A. V\'azquez, Gen. Rel. Grav. {\bf 43}, 1153, (2011). 

\bibitem{quev08}
H. Quevedo and A. S\'anchez, JHEP {\bf 09}, 034, (2008).

\bibitem{KSBH} W. Janke, D. A. Johnston and R. Kenna, J. Phys. A {\bf 43}, 425206, (2010).

\bibitem{quevdiego} H. Quevedo and D. Tapias, arXiv:1301.0262.

\bibitem{HanChen}
Y. Han and G. Chen, Phys. Lett. B {\bf 714}, 127-130,  (2012).

\bibitem{AlejQuev}
A. Aviles, A. Bastarrachea-Almodovar, L. Campuzano and H. Quevedo,  Phys. Rev. D {\bf 86}, 063508, (2012).

\bibitem{Brasil}
M. E. Rodrigues and Z. A. A. Oporto, Phys. Rev. D {\bf 85}, 104022, (2012).




\bibitem{Termometrica}
A. Bravetti and F. Nettel, arXiv:1208.0399.



\bibitem{AleQuevDavood}
A. Bravetti, D. Momeni, R. Myrzakulov and H. Quevedo, {\it Geometrothermodynamics of higher dimensional black holes}, to appear in Gen. Rel. Grav., doi:10.1007/s10714-013-1549-2 .



\bibitem{aman2bis}
J. E. \AA man and N. Pidokrajt, arXiv:1004.5550.

\bibitem{EmpMyers}
R. Emparan and R. C. Myers, JHEP  {\bf 0309}, 025, (2003).

\bibitem{MonteiroPerry}
R. Monteiro, M. J. Perry and J. E. Santos, Phys. Rev. D {\bf 80}, 024041,  (2009).

\bibitem{AsteMann}
D. Astefanesei, R. B. Mann, M. J. Rodriguez and C. Stelea, Class. Quant. Grav. {\bf 27}, 165004,  (2010).

\bibitem{AsteRodriguez}
D. Astefanesei, M. J. Rodriguez and S. Theisen, JHEP {\bf 1008}, 046, (2010).    


\bibitem{arnold} V. I. Arnold, {\it Mathematical Methods of Classical Mechanics}
(Springer Verlag, New York, 1980).




\bibitem{Callen}
H. B. Callen, {\it   Thermodynamics and an Introduction to Thermostatics}
(John Wiley and Sons, Inc., New York, 1985).



\bibitem{NewDev}
A. Bravetti, C. S. Lopez-Monsalvo, F. Nettel and H. Quevedo, J. Math. Phys. {\bf 54}, 033513, (2013).



\bibitem{MPoriginal}
R. C. Myers and M. J. Perry, Annals Phys. {\bf 172}, 304, (1986).


\bibitem{ERBR}
R. Emparan and H. S. Reall, Living Rev. Rel. {\bf 11}, 6, (2008).


\bibitem{Chamb1}
A. Chamblin, R. Emparan, C. V. Johnson and R. C. Myers, Phys. Rev. D {\bf 60}, 064018, (1999).  

\bibitem{Chamb2}
A. Chamblin, R. Emparan, C. V. Johnson and R. C. Myers, Phys. Rev. D {\bf 60}, 104026, (1999).  





\end{thebibliography}
\end{document}